\documentclass[12pt]{article}

\usepackage[utf8]{inputenc}
\usepackage[T1]{fontenc}
\usepackage{amsmath,amssymb}
\usepackage{graphicx}
\usepackage{booktabs}
\usepackage{natbib}
\usepackage{hyperref}
\usepackage[margin=2.5cm]{geometry}
\usepackage{setspace}
\usepackage{caption}
\usepackage{ragged2e}
\usepackage{subcaption}
\usepackage{float}
\usepackage{pgfplots}
\usepackage{tikz}
\pgfplotsset{
  compat=1.18,
  /pgf/number format/1000 sep={}
}
\usepgfplotslibrary{statistics}

\title{Public sentiments on the fourth industrial
revolution:
An unsolicited public opinion poll from Twitter}

\author{Diletta Abbonato\thanks{CPS, University of Turin Email: diletta.abbonato@unito.it }}

\date{\today}

\begin{document}

\maketitle

\begin{abstract}
This paper establishes an empirical baseline of public sentiment toward Fourth Industrial Revolution (4IR) technologies across six European countries during the period 2006--2019, prior to the widespread adoption of generative AI systems. Employing transformer-based natural language processing models on a corpus of approximately 90,000 tweets and news articles, I document a European public sphere increasingly divided in its assessment of technological change: neutral sentiment declined markedly over the study period as citizens sorted into camps of enthusiasm and concern, a pattern that manifests distinctively across national contexts and technology domains. Approximately 6\% of users inhabit echo chambers characterized by sentiment-aligned networks, with privacy discourse exhibiting the highest susceptibility to such dynamics. These findings provide a methodologically rigorous reference point for evaluating how the introduction of ChatGPT and subsequent generative AI systems has transformed public discourse on automation, employment, and technological change. The results carry implications for policymakers seeking to align technological governance with societal values in an era of rapid AI advancement.

\end{abstract}

\newpage

\section{Introduction}

The relationship between technological change and public sentiment constitutes a fundamental concern for both economic policy and democratic governance. As data-driven technologies increasingly permeate societal processes, understanding how citizens perceive, interpret, and respond to these transformations becomes essential for designing regulatory frameworks that command public legitimacy \citep{douglas2012social}. This imperative has acquired particular urgency following the release of ChatGPT in November 2022, an event that dramatically altered public awareness of artificial intelligence capabilities and catalyzed widespread debate on automation, employment, and the future of human-machine interaction.

Yet scholarly understanding of how public sentiment toward AI and related technologies evolved \textit{prior} to this inflection point remains surprisingly limited. While a substantial literature examines technology acceptance at the individual level \citep{rogers1962diffusion, hekkert2007functions} and investigates media representations of AI \citep{fast2017long, cave2019hopes}, large-scale empirical analyses of public discourse across multiple technologies and national contexts are scarce. This gap impedes the capacity to assess whether---and how---generative AI has genuinely transformed public attitudes, as opposed to merely amplifying pre-existing patterns of enthusiasm, skepticism, or polarization.

 Drawing on approximately 90,000 tweets and news articles discussing artificial intelligence, robotics, blockchain, cloud computing, the Internet of Things, and virtual reality, I employ transformer-based natural language processing models to systematically characterize sentiment dynamics prior to the generative AI era. Social media discourse, while not representative of general public opinion, provides a valuable window into how technology narratives form and circulate among engaged citizens---the population most likely to shape and respond to policy debates.
My analysis documents patterns of polarization, identifies thematic concerns, and examines evidence of echo chamber formation in technology-related discourse.

The Fourth Industrial Revolution framework, as articulated by \citet{schwab2017fourth}, provides a conceptually coherent lens for this investigation. Unlike analyses focused narrowly on artificial intelligence, the 4IR perspective recognizes that public perceptions are shaped by the broader constellation of technologies---automation, connectivity, decentralization---that collectively restructure economic and social relations \citep{geels2002technological, brynjolfsson2014second}. Citizens do not encounter AI in isolation; they experience it alongside smart devices, algorithmic platforms, and automated services that together constitute the technological substrate of contemporary life.

The contribution is threefold. First, I provide what is, to myknowledge, the first large-scale comparative analysis of public sentiment toward 4IR technologies across multiple European countries, documenting substantial cross-national variation that reflects differing institutional contexts and technology governance regimes. Second, I demonstrate the utility of multilingual transformer models---specifically XLM-RoBERTa for sentiment classification and DeBERTa for thematic categorization---as tools for social science research on public discourse, contributing to the growing methodological literature on computational text analysis in economics \citep{gentzkow2019text}. Third, and most importantly, I establish a pre-ChatGPT baseline against which subsequent shifts in public sentiment can be rigorously evaluated.

This final contribution merits emphasis. The release of ChatGPT represented what economic historians term a ``focusing event''---a development that concentrates public attention and potentially reshapes collective understanding. Assessing its impact requires counterfactual reasoning: what would public sentiment look like absent this intervention? The analysis provides the empirical foundation for such assessments by documenting the trajectory of sentiment prior to November 2022. Researchers investigating post-ChatGPT discourse can employ the findings as a reference point, distinguishing genuine attitude shifts from continuations of pre-existing trends.

The results reveal a public discourse characterized by increasing polarization. Over the 2006--2019 period, neutral sentiment declined markedly in both media coverage and user responses, while positive and negative expressions intensified. This pattern---observed across technologies and national contexts---suggests that engagement with 4IR technologies progressively sorted citizens into camps of enthusiasm and concern, eroding the middle ground of ambivalence or indifference. Thematic analysis indicates that employment implications and environmental considerations dominated public discussion, while privacy concerns, though less frequent, exhibited the highest susceptibility to misinformation.

These findings carry direct implications for technology governance. Policymakers confronting the challenges of AI regulation operate within a public sphere already structured by years of debate, narrative formation, and community polarization. Effective governance requires understanding this pre-existing terrain---the hopes and fears that citizens bring to encounters with new technologies, the information environments in which they form opinions, and the social dynamics that amplify or attenuate particular viewpoints. The analysis highlights this terrain, providing an evidence base for regulatory design that acknowledges the complexity of public sentiment toward technological change.

The remainder of the paper proceeds as follows. Section~\ref{sec:theory} develops the theoretical framework, integrating literature on technology narratives, public opinion formation, and the economics of information environments. Section~\ref{sec:methods} describes the  methodological approach, with particular attention to the transformer architectures employed for sentiment and thematic classification. Section~\ref{sec:data} presents the data construction process and descriptive statistics. Section~\ref{sec:results} reports findings on sentiment dynamics, cross-national variation, and echo chamber identification. Section~\ref{sec:discussion} interprets these results in light of subsequent developments in generative AI and draws implications for policy and future research.

\section{Theoretical Framework}
\label{sec:theory}

Understanding public sentiment toward emerging technologies requires integrating insights from multiple disciplinary traditions. I draw on three interconnected bodies of literature: (i) research on technology narratives and their societal functions; (ii) economic theories of innovation diffusion and technology acceptance; and (iii) studies of information environments, with particular attention to social media dynamics. Together, these perspectives illuminate why public attitudes toward 4IR technologies exhibit the patterns of polarization, thematic clustering, and community segmentation that the empirical analysis documents.

\subsection{Technology Narratives: Hopes and Fears}
\label{subsec:narratives}

Technologies do not enter public consciousness as neutral artifacts; they arrive embedded in narratives that shape interpretation, emotional response, and behavioral intention. Research in science and technology studies has long emphasized that public understanding of technology is mediated by culturally available stories, metaphors, and imaginaries \citep{jasanoff2015dreamscapes, latour2007reassembling}. These narratives perform crucial cognitive and social functions: they render complex technologies comprehensible, provide frameworks for evaluating risks and benefits, and orient collective action toward technological futures \citep{suchman2007human}.

\citet{cave2019hopes} provide a systematic taxonomy of AI narratives organized around four fundamental tensions, each reflecting deep-seated hopes and corresponding fears (Table~\ref{tab:narratives}). The \textit{Immortality--Dehumanization} axis captures narratives in which AI promises to transcend biological limitations---conquering disease, extending lifespan, augmenting cognition---while simultaneously threatening the essence of human identity and value. The \textit{Freedom--Obsolescence} dimension addresses labor market implications: optimistic visions of liberation from tedious or dangerous work contrast with anxieties about technological unemployment and economic displacement. The \textit{Gratification--Alienation} tension concerns social relations, juxtaposing scenarios of AI-enabled connection and fulfillment against fears of atomization and the replacement of human intimacy with machine interaction. Finally, the \textit{Dominance--Uprising} axis encompasses narratives of security and control, from AI-enhanced protection against threats to science fiction scenarios of machine rebellion.

\begin{table}[htbp]
\centering
\caption{Narrative Tensions in Public Discourse on AI and 4IR Technologies}
\label{tab:narratives}
\begin{tabular}{llll}
\toprule
\textbf{Domain} & \textbf{Hope} & \textbf{Fear} & \textbf{Core Tension} \\
\midrule
Health/Biology & Immortality & Dehumanization & Human enhancement vs. loss of essence \\
Employment & Freedom & Obsolescence & Liberation from labor vs. displacement \\
Social Relations & Gratification & Alienation & Connection vs. isolation \\
Security/Control & Dominance & Uprising & Protection vs. loss of control \\
\bottomrule
\end{tabular}
\smallskip
\begin{minipage}{0.9\textwidth}
\footnotesize\textit{Note:} Framework adapted from \citet{cave2019hopes}. Each dimension represents a continuum along which public discourse positions specific technologies and applications.
\end{minipage}
\end{table}

These narrative tensions are not confined to artificial intelligence narrowly construed. The broader suite of 4IR technologies activates analogous imaginaries. Blockchain evokes narratives of decentralization and trust---freedom from institutional intermediaries---alongside concerns about complexity, energy consumption, and facilitation of illicit activity \citep{khan2018iot}. Virtual reality promises immersive experiences that transcend physical limitations while raising anxieties about escapism and detachment from embodied social life. The Internet of Things offers visions of seamless efficiency and connectivity counterbalanced by fears of surveillance and security vulnerabilities \citep{weinberg2015internet, atzori2010internet}.

Crucially, these narratives exhibit temporal dynamics. \citet{fast2017long} document systematic shifts in media coverage of AI over several decades, finding oscillation between optimistic and pessimistic framings that correlate with technological developments, economic conditions, and cultural events. Public sentiment, in this view, is not static but evolves in response to both technological change and the narrative frameworks available for interpreting that change. The empirical analysis captures one such evolutionary trajectory---the period during which 4IR technologies transitioned from specialist concerns to mainstream public awareness, but before generative AI dramatically recentered the discourse.

\subsection{Economic Perspectives on Technology Adoption}
\label{subsec:economics}

Economic analysis contributes two essential perspectives to understanding public sentiment toward emerging technologies: theories of innovation diffusion that explain adoption patterns, and public economics frameworks that illuminate collective decision-making about technological futures.

The diffusion of innovations paradigm, originating with \citet{rogers1962diffusion}, models technology adoption as a social process in which information flows through networks, early adopters signal quality and appropriateness, and adoption decisions reflect both individual cost-benefit calculations and social influence. This framework explains the characteristic S-curve of technology penetration and the segmentation of populations into adopter categories---innovators, early adopters, early majority, late majority, and laggards---each characterized by distinctive attitudes toward novelty and risk.

For understanding public sentiment, diffusion theory suggests that attitudes should evolve predictably as technologies mature. Early discourse, dominated by innovators and technology enthusiasts, tends toward optimism; as technologies diffuse to broader populations with more heterogeneous preferences and risk tolerances, sentiment diversifies and often becomes more critical \citep{greenhalgh2017beyond}. The democratization of digital technologies---smartphones, social media, algorithmic services---has accelerated this process, compressing the timeline over which technologies move from elite novelty to mass experience \citep{ceruzzi2012computing, west2012digital}.

Yet diffusion theory, focused primarily on adoption decisions, incompletely captures the formation of attitudes toward technologies that citizens encounter whether or not they choose to adopt them. Here, public economics perspectives become essential. Technologies like AI and automation generate externalities---effects on employment, privacy, social relations, and power distributions that extend beyond direct users to affect society broadly \citep{acemoglu2020robots, autor2015there}. Public sentiment toward such technologies reflects not merely individual adoption calculus but collective judgments about societal impacts and distributional consequences.

\citet{brynjolfsson2014second} characterize the contemporary technological moment as one of ``bounty and spread''---unprecedented productive potential accompanied by increasingly unequal distribution of benefits. Public awareness of this dynamic shapes sentiment: enthusiasm for technological capability coexists with anxiety about who gains and who loses. Survey research consistently documents this ambivalence. \citet{zhang2020public} find that American respondents simultaneously express optimism about AI's potential to improve healthcare and scientific research while voicing concern about privacy, accountability, and employment effects. \citet{kelley2021exciting} report similar patterns across eight countries, with support for AI applications varying systematically by domain---highest for healthcare and science, lowest for surveillance and autonomous weapons.

This domain-specificity of attitudes has important implications for analyzing public discourse. Sentiment toward ``AI'' or ``robots'' in the abstract may be less meaningful than sentiment toward specific applications in specific contexts. The thematic classification approach, employing transformer-based models to categorize discourse by domain (employment, environment, health, privacy), captures this specificity, enabling analysis of how sentiment varies across the application areas that matter most for policy.

\subsection{Information Environments and Opinion Dynamics}
\label{subsec:information}

Public sentiment forms within information environments that powerfully shape what citizens know, believe, and feel about technological change. The transformation of these environments by social media platforms constitutes perhaps the most significant development in public opinion formation over the period the data cover.

Social networks have become primary channels through which citizens encounter information about emerging technologies \citep{allcott2017social}. Platforms like Twitter enable rapid dissemination of news, commentary, and opinion, dramatically expanding the range of voices participating in technology discourse beyond traditional media gatekeepers. This democratization carries both promise and peril. On one hand, it enables broader participation in debates previously confined to experts and elites; on the other, it creates conditions favorable to misinformation, polarization, and fragmentation of the public sphere \citep{lazer2018science, vosoughi2018spread}.

Three features of social media environments are particularly relevant for understanding technology discourse. First, \textit{algorithmic curation} shapes exposure. Platform algorithms, optimizing for engagement, tend to surface content that provokes strong emotional responses---a dynamic that may systematically amplify extreme positions while suppressing nuanced or moderate viewpoints \citep{pariser2011filter, gillespie2018custodians}. If polarization in technology sentiment has increased over time, algorithmic amplification represents a plausible mechanism.

Second, \textit{network homophily} and \textit{selective exposure} can generate echo chambers---information environments in which users predominantly encounter viewpoints consonant with their existing beliefs \citep{sunstein2017republic}. \citet{del2016spreading} document that on social media, misinformation and factual content spread through largely distinct networks, with users clustering into communities defined by shared epistemological orientations. For technology discourse, this implies that enthusiasts and skeptics may inhabit increasingly separate information worlds, reinforcing rather than challenging their respective positions.

Third, \textit{misinformation} circulates readily in social media environments, particularly on topics characterized by complexity, uncertainty, and high emotional stakes \citep{lewandowsky2017beyond}. Emerging technologies exemplify these conditions. Claims about AI capabilities, automation's employment effects, or blockchain's transformative potential are difficult for non-experts to evaluate, creating vulnerability to misleading narratives. \citet{mosleh2022measuring} develop methods for quantifying exposure to misinformation based on the information diets implied by social network connections, an approach I adapt to assess misinformation exposure in technology discourse.

The theoretical expectation, combining these elements, is that social media discourse on 4IR technologies should exhibit (i) increasing polarization as neutral positions erode under pressure from engagement-optimizing algorithms; (ii) community segmentation as homophily and selective exposure sort users into like-minded clusters; and (iii) differential susceptibility to misinformation across technology domains and user communities. The empirical analysis tests these expectations against data from the pre-generative AI period, establishing baseline patterns against which post-ChatGPT dynamics can be compared.

\begin{figure}[H]
    \caption{Different aspects of public opinion dynamics}
    \centering
    \begin{subfigure}{0.45\textwidth}
        \centering
        \includegraphics[width=\linewidth]{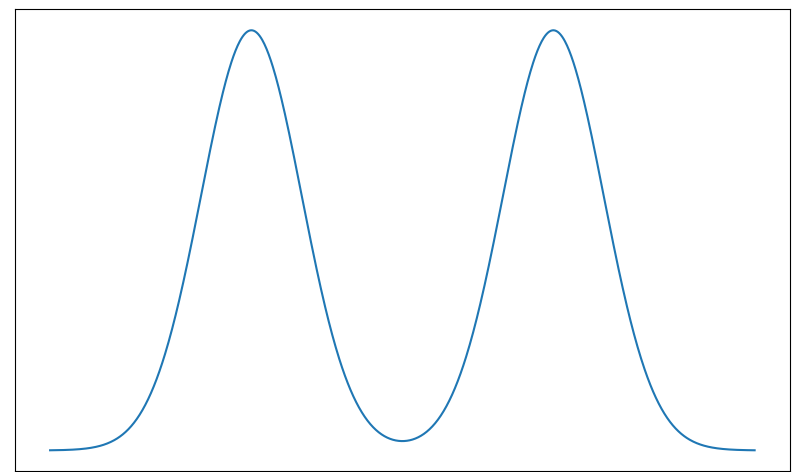}
        \caption{Polarization}
        \label{fig:Polarization}
    \end{subfigure}
    \hfill
    \begin{subfigure}{0.45\textwidth}
        \centering
        \includegraphics[width=\linewidth]{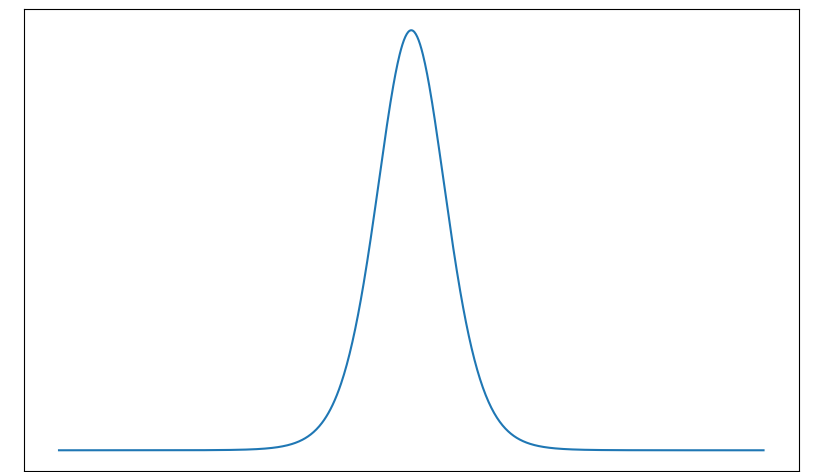}
        \caption{Consensus}
        \label{fig:Consensus}
    \end{subfigure}
    \hfill
    \begin{subfigure}{0.45\textwidth}
        \centering
        \includegraphics[width=\linewidth]{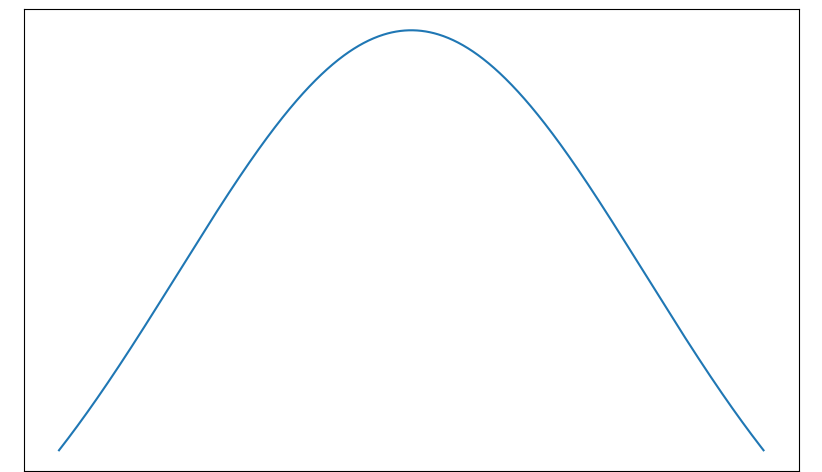}
        \caption{Dissent}
        \label{fig:Dissent}
    \end{subfigure}
    \\
    \small 
\justifying
\textit{Notes}: In figure (a) population is divided into two dominant groups with opposing views on a specific issue. The peaks indicate the concentration of individuals within each opinion group, while the trough indicates a lack of moderate stances. This highlights the clear divide and potential for increased social tensions; in figure (b) the ``lock-in" effect in public opinion occurs when a single viewpoint has become overwhelmingly predominant. This marginalizes alternative perspectives and demonstrates the societal or cultural homogeneity on a specific issue; in figure (c) the dissent shows a spectrum of views where the majority holds a central opinion, while a range of dissenting views exists on either side. This indicates a diverse and engaged public discourse
    \label{fig:main}
\end{figure}

\subsection{Integrating Perspectives: A Framework for Analysis}
\label{subsec:integration}

The foregoing perspectives converge on a conception of public sentiment toward 4IR technologies as a dynamic, socially embedded phenomenon shaped by narrative availability, economic stakes, and information environment characteristics. Figure~\ref{fig:framework} schematizes these relationships.

\begin{figure}[htbp]
\centering
\fbox{\parbox{0.85\textwidth}{
\centering
\textbf{Analytical Framework}\\[0.5em]
\small
\textbf{Technology Narratives} (Hopes/Fears, Cultural Imaginaries)\\
$\downarrow$\\
\textbf{Information Environment} (Social Media, Algorithmic Curation, Network Structure)\\
$\downarrow$\\
\textbf{Public Sentiment} (Valence, Intensity, Thematic Focus)\\
$\downarrow$\\
\textbf{Observable Patterns} (Polarization, Echo Chambers, Cross-National Variation)
}}
\caption{Conceptual framework for analyzing public sentiment toward 4IR technologies}
\label{fig:framework}
\end{figure}

Citizens interpret technologies through available narratives, which provide cognitive schemas for understanding novel phenomena. These interpretations are shaped by---and expressed within---information environments that filter, amplify, and sort content according to platform logics and network structures. The resulting sentiment exhibits measurable patterns: aggregate distributions of positive, negative, and neutral expressions; clustering by thematic concern; and community structures reflecting homophily in attitudes.

This framework generates specific empirical expectations. I anticipate declining neutrality as engagement with 4IR technologies matures and citizens form more definite opinions. I expect sentiment to vary by technology domain, reflecting the distinct narrative tensions each technology activates. I predict cross-national variation corresponding to differences in technology governance regimes, labor market structures, and media environments. And I expect evidence of echo chamber formation, with users clustering into communities of shared sentiment.

The following sections describe the methodological approach to testing these expectations and present the resulting empirical findings.

\section{Methodological Approach}
\label{sec:methods}

Analyzing public sentiment at scale requires computational methods capable of processing large volumes of multilingual text while capturing the nuanced valence and thematic content of social media discourse. This section describes the methodological approach, focusing on the transformer-based natural language processing models employed for sentiment classification and thematic categorization. I present these methods as tools for social science inquiry rather than as contributions to machine learning per se, emphasizing their suitability for the substantive research questions motivating this study.

\subsection{Transformer Models for Text Analysis}
\label{subsec:transformers}

The analysis of textual data has been revolutionized by transformer architectures, a class of neural network models that achieve state-of-the-art performance on a wide range of natural language processing tasks \citep{vaswani2017attention}. Unlike earlier approaches that processed text sequentially, transformers employ an ``attention'' mechanism that enables simultaneous consideration of all words in a text, capturing long-range dependencies and contextual relationships that sequential models often miss.

For social science applications, transformers offer several advantages over traditional methods such as dictionary-based sentiment analysis or bag-of-words classifiers. First, they handle \textit{context-dependent meaning}: the same word can carry different sentiment depending on surrounding text, and transformers capture these contextual effects. Second, they manage \textit{negation and qualification}: phrases like ``not good'' or ``somewhat concerning'' that confound simpler methods are processed appropriately. Third, contemporary transformer models are \textit{pre-trained} on massive text corpora, encoding broad linguistic knowledge that can be applied to specific classification tasks with relatively modest additional training \citep{devlin2019bert}.

The analysis employs two transformer-based models, each suited to a specific analytical task: XLM-RoBERTa for sentiment classification and DeBERTa for thematic categorization.

\subsection{Sentiment Classification with XLM-RoBERTa}
\label{subsec:xlmroberta}

Sentiment analysis---the classification of text according to expressed positive, negative, or neutral sentiment---constitutes the core empirical task of this study. I employ XLM-RoBERTa (Cross-lingual Language Model - Robustly Optimized BERT Pretraining Approach), a multilingual transformer model developed by \citet{conneau2020unsupervised} and fine-tuned for Twitter sentiment analysis by \citet{barbieri2021xlm}.

The choice of XLM-RoBERTa reflects the multilingual nature of the corpus, which spans six European languages (English, French, German, Italian, Spanish, and Dutch). Traditional sentiment analysis approaches require separate models or lexicons for each language, introducing potential inconsistencies in cross-national comparisons. XLM-RoBERTa, by contrast, is trained on text from 100 languages simultaneously, learning cross-lingual representations that enable consistent sentiment classification regardless of source language. This architecture is particularly valuable for comparative research, as it ensures that observed cross-national differences reflect genuine variation in sentiment rather than artifacts of language-specific model performance.

The specific model variant I employ (XLM-T) was fine-tuned on a corpus of over 198 million tweets across multiple languages, with supervised training on sentiment-labeled datasets \citep{barbieri2021xlm}. This Twitter-specific training is essential for the application: social media text exhibits distinctive characteristics---informal language, abbreviations, emoji, hashtags---that models trained on formal text corpora handle poorly. XLM-T achieves state-of-the-art performance on multilingual Twitter sentiment benchmarks, with macro-averaged F1 scores exceeding 0.69 across eight languages.

For each text in the corpus, the model produces probability distributions over three sentiment classes (positive, negative, neutral). I assign each text to the highest-probability class for categorical analyses while retaining continuous probability scores for analyses requiring finer gradation. This approach enables both aggregate characterization of sentiment distributions and examination of sentiment intensity.

\subsection{Thematic Classification with Zero-Shot Learning}
\label{subsec:deberta}

Beyond sentiment valence, understanding public discourse requires identifying the thematic concerns that structure debate. What aspects of 4IR technologies attract attention? Do discussions focus on employment implications, privacy risks, health applications, or environmental considerations? Answering these questions requires classifying texts by topic---a task complicated by the absence of pre-labeled training data for the specific thematic categories.

I address this challenge using zero-shot classification, a machine learning paradigm in which models classify texts into categories not encountered during training \citep{yin2019benchmarking}. Zero-shot classification leverages the semantic knowledge encoded in large language models: if a model understands what ``employment'' and ``privacy'' mean, it can assess whether a given text is more likely to concern one topic or the other, even without explicit training examples.

I employ DeBERTa (Decoding-enhanced BERT with disentangled attention), a transformer architecture that achieves strong performance on natural language inference tasks underlying zero-shot classification \citep{he2021deberta}. DeBERTa's key innovation is a disentangled attention mechanism that separately encodes content and position information, improving the model's ability to capture subtle semantic relationships. On the MultiNLI benchmark for natural language inference, DeBERTa achieves accuracy of 91.3\%, among the highest reported for models of comparable scale.

The thematic classification employs five categories derived from the theoretical framework: \textit{employment} (labor market implications, automation, job displacement), \textit{environment} (sustainability, energy consumption, climate), \textit{health} (medical applications, well-being, biological enhancement), \textit{privacy} (data protection, surveillance, consent), and \textit{other} (residual category for texts not clearly aligned with the preceding themes). These categories map onto the narrative tensions identified by \citet{cave2019hopes} while reflecting the specific policy domains most salient in 4IR discourse.

For each text, DeBERTa produces probability scores for each thematic category. I assign texts to the highest-probability category, yielding a thematic distribution that characterizes the substantive focus of public discourse. This approach enables analysis of how thematic attention varies across technologies, countries, and time periods.

\subsection{Echo Chamber Identification}
\label{subsec:echochamber}

The theoretical framework predicts that social media discourse may exhibit echo chamber dynamics, with users clustering into communities of shared sentiment. Testing this prediction requires operationalizing the echo chamber concept and developing methods for its empirical detection.

I define a user as situated within an echo chamber when their expressed sentiment aligns closely with the sentiment expressed by their network connections (followers and accounts followed). This operationalization captures the core intuition that echo chambers involve reinforcement of existing viewpoints through selective exposure to like-minded others \citep{sunstein2017republic, quattrociocchi2016echo}.

Specifically, for each user $i$, I compute the average sentiment of tweets by accounts in their follower network and following network. Let $\bar{s}_i^{\text{followers}}$ and $\bar{s}_i^{\text{following}}$ denote these averages, where sentiment is coded numerically (positive = 1, neutral = 0.5, negative = 0). I classify user $i$ as situated within an echo chamber if both averages exceed a threshold $\tau = 0.6$ (indicating predominantly positive network sentiment aligned with user's positive sentiment) or both fall below $1 - \tau = 0.4$ (indicating predominantly negative aligned sentiment).

The threshold value of 0.6 balances sensitivity and specificity. Lower thresholds would classify weakly aligned users as echo chamber participants, potentially overstating the phenomenon; higher thresholds would require near-unanimity, potentially understating it. The choice aligns with approaches in the literature \citep{cinelli2021echo} and proves robust to moderate variation in sensitivity analyses.

\subsection{Misinformation Exposure Assessment}
\label{subsec:misinfo}

To assess the relationship between technology discourse and misinformation, I employ the elite misinformation-exposure score developed by \citet{mosleh2022measuring}. This measure leverages the insight that users' information diets can be inferred from the accounts they follow: following sources known to disseminate low-quality or false information implies exposure to such content.

\citet{mosleh2022measuring} construct their measure by identifying Twitter accounts of political elites and media outlets, then scoring each based on the quality of information they typically share (assessed through fact-checking databases and expert ratings). A user's misinformation-exposure score is computed as the average score of elite accounts in their following network, weighted by the frequency of those accounts' tweeting activity.

I apply this measure to users in the corpus, obtaining misinformation-exposure scores that I then relate to sentiment patterns and echo chamber participation. This enables analysis of whether users with higher misinformation exposure exhibit distinctive sentiment profiles or greater propensity to inhabit echo chambers---relationships with important implications for understanding the information environments surrounding technology discourse.

\subsection{Methodological Considerations}
\label{subsec:limitations}

Several features of the methodological approach define the scope of valid inference from the findings.

First, the sampling strategy targets discourse among Twitter users who engage with mainstream media coverage of 4IR technologies. This design choice reflects a deliberate focus on the intersection between media framing and public response---a theoretically motivated population for understanding how technology narratives circulate and evolve. The resulting findings characterize sentiment dynamics within this engaged public rather than the general population, a distinction common to social media research \citep{ruths2014social}.

Second, transformer-based classifiers, while substantially outperforming traditional methods, introduce measurement error. I address this by focusing on aggregate patterns and temporal trends, where random classification errors attenuate. The consistency of declining neutrality across technologies, countries, and thematic domains---each representing independent classification decisions---provides internal validation that the core finding is robust to plausible error rates.

Third, the echo chamber operationalization based on network sentiment alignment captures one empirically tractable dimension of a multifaceted phenomenon. The 6\% prevalence estimate should be understood as identifying users with strong, observable sentiment homophily rather than exhaustively characterizing epistemically constrained information environments.

Finally, the analysis is descriptive rather than causal. I document systematic patterns in sentiment dynamics without claiming to identify the mechanisms generating them. This descriptive contribution is appropriate for establishing the baseline that motivates future causal inquiry.

\section{Data and Empirical Strategy}
\label{sec:data}

This section describes the construction of the dataset, the sampling strategy employed to identify relevant discourse, and descriptive statistics characterizing the resulting corpus. The data collection process involved multiple stages: identification of media sources, keyword-based filtering for 4IR-relevant content, retrieval of user interactions, and network expansion to capture follower and following relationships.

\subsection{Sampling Strategy and Data Sources}
\label{subsec:sampling}

The empirical strategy targets technology discourse as mediated by major newspapers and the users who engage with their coverage. This approach offers several advantages over alternative sampling methods. Random sampling of Twitter users would yield a corpus dominated by content unrelated to technology; keyword-based sampling of all tweets would capture enormous volumes but with uncertain provenance and quality. By anchoring the sample in newspaper coverage, I capture discourse that responds to---and extends---mainstream media framing of 4IR technologies, enabling analysis of the interplay between media narratives and public response.

I focus on six European countries representing diverse linguistic, economic, and institutional contexts: France, Germany, Italy, the Netherlands, Spain, and the United Kingdom. These countries span Northern and Southern Europe, include both Anglophone and non-Anglophone contexts, and exhibit variation in technology sector development, labor market institutions, and regulatory approaches to digital technologies. This diversity enables examination of cross-national variation while maintaining feasibility in data collection and linguistic processing.

For each country, I identified the most widely circulated newspapers based on audited circulation figures (Table~\ref{tab:newspapers}). I then retrieved the official Twitter accounts of these newspapers and collected all tweets from these accounts containing keywords related to 4IR technologies. Keywords were specified in both English and the relevant national language to ensure comprehensive coverage (Table~\ref{tab:keywords}).

\begin{table}[htbp]
\centering
\caption{Newspapers and Twitter Accounts by Country}
\label{tab:newspapers}
\small
\begin{tabular}{ll}
\toprule
\textbf{Country} & \textbf{Newspaper Twitter Accounts} \\
\midrule
United Kingdom & DailyMailUK, guardiannews, TheSun, thetimes, DailyMirror \\
France & laborede, Le\_Figaro, LesEchos, le\_Parisien, Mediapart \\
Germany & BILD, faznet, SZ, welt, handelsblatt, taboresspiegel \\
Italy & repubblica, Corriere, LaStampa, Sole24ore, fattoquotidiano \\
the Netherlands & Telegraaf, volkskrant, nrc, trouw, ADnl \\
Spain & el\_pais, abc\_es, LaVanguardia, elperiodico, elmaboroEs \\
\bottomrule
\end{tabular}
\end{table}

\begin{table}[htbp]
\centering
\caption{4IR Technology Keywords by Language}
\label{tab:keywords}
\small
\begin{tabular}{ll}
\toprule
\textbf{Technology} & \textbf{Keywords (English + Translations)} \\
\midrule
Artificial Intelligence & artificial intelligence, AI, [national language equivalents] \\
Robotics & robot, robotics, automation \\
Blockchain & blockchain, cryptocurrency, bitcoin \\
Cloud Computing & cloud computing, cloud services \\
Internet of Things & IoT, internet of things, smart devices \\
Virtual Reality & virtual reality, VR, augmented reality \\
5G & 5G, fifth generation network \\
\bottomrule
\end{tabular}
\smallskip
\begin{minipage}{0.9\textwidth}
\footnotesize\textit{Note:} Keywords were translated into French, German, Italian, Dutch, and Spanish with attention to linguistic precision and cultural appropriateness. For example, Italian searches included both ``artificial intelligence'' and ``intelligenza artificiale.''
\end{minipage}
\end{table}

\subsection{User Sampling and Network Expansion}
\label{subsec:usersampling}

From the newspaper tweets identified through keyword filtering, I constructed the user sample by collecting all users who interacted with these tweets through likes, retweets, or comments. This interaction-based sampling identifies users who demonstrated active engagement with 4IR-related media coverage, providing a sample of individuals with revealed interest in technology discourse.

For each user in this initial sample, I retrieved their complete tweet timeline for the period January 2006 through December 2019 using the Twitter API via the twarc2 library. This fourteen-year window captures the emergence and maturation of public discourse on 4IR technologies while ending before two potentially confounding events: the COVID-19 pandemic (which dramatically shifted public attention and technology usage patterns) and the release of ChatGPT (which transformed AI discourse specifically).

To enable echo chamber analysis, I additionally collected data on the follower and following networks of sampled users. For a randomly selected subset of users, I retrieved the accounts they follow and the accounts following them, then collected tweets from these network connections. This expansion enables assessment of sentiment alignment between users and their network contacts.

\subsection{Corpus Characteristics}
\label{subsec:corpus}

The final dataset comprises approximately 25,000 unique users and 90,000 tweets. Table~\ref{tab:corpus} presents the distribution of observations across countries, distinguishing between newspaper tweets (media coverage) and user tweets (public response).

\begin{table}[htbp]
\centering
\caption{Corpus Size by Country}
\label{tab:corpus}
\begin{tabular}{lrrr}
\toprule
\textbf{Country} & \textbf{News Tweets} & \textbf{User Tweets} & \textbf{Total} \\
\midrule
United Kingdom & 5,922 & 11,651 & 17,573 \\
France & 5,491 & 9,803 & 15,294 \\
Italy & 3,705 & 4,728 & 8,433 \\
Spain & 3,585 & 4,804 & 8,389 \\
Germany & 852 & 1,375 & 2,227 \\
the Netherlands & 1,087 & 1,209 & 2,296 \\
\midrule
\textbf{Total} & 20,642 & 33,570 & 54,212 \\
\bottomrule
\end{tabular}
\smallskip
\begin{minipage}{0.9\textwidth}
\footnotesize\textit{Note:} Counts reflect tweets containing 4IR keywords after preprocessing and deduplication. The full corpus including network-expanded users comprises approximately 90,000 tweets.
\end{minipage}
\end{table}

The United Kingdom and France contribute the largest shares of observations, reflecting both the size of their Twitter-using populations and the volume of English and French technology coverage. Germany and the Netherlands contribute smaller shares, partly reflecting lower Twitter penetration in these countries during the study period. These imbalances are addressed in the analysis through country-specific examination rather than pooled analysis that would weight observations by corpus size.

\subsection{Technology and Thematic Distributions}
\label{subsec:distributions}

Figure~\ref{fig:keywords} presents the distribution of tweets across 4IR technology categories. Robotics discourse dominates the corpus (approximately 35,000 mentions), followed by blockchain (27,000) and artificial intelligence (16,000). Cloud computing and IoT receive substantially less attention, consistent with their lower public salience relative to more visible technologies like robots and AI.

\begin{figure}[htbp]
\centering
\begin{tikzpicture}
\begin{axis}[
    ybar,
    width=0.9\textwidth,
    height=7cm,
    ylabel={Number of Mentions},
    ylabel style={font=\small},
    symbolic x coords={AI, VR, Blockchain, Robot, Cloud, IoT, 5G},
    xtick=data,
    xticklabel style={font=\small},
    yticklabel style={font=\small},
    nodes near coords,
    nodes near coords style={font=\scriptsize, above},
    ymin=0,
    ymax=40000,
    bar width=0.6cm,
    fill=blue!60,
    every node near coord/.append style={rotate=0, anchor=south},
]
\addplot[fill=blue!60, draw=blue!80] coordinates {
    (AI, 16000)
    (VR, 4000)
    (Blockchain, 27000)
    (Robot, 35000)
    (Cloud, 1000)
    (IoT, 4000)
    (5G, 3500)
};
\end{axis}
\end{tikzpicture}
\caption{Frequency of 4IR technology mentions in corpus}
\label{fig:keywords}
\end{figure}

Table~\ref{tab:techbycountry} disaggregates technology attention by country, revealing notable cross-national variation. Blockchain discourse is particularly prominent in Germany (49.7\% of technology mentions) and the United Kingdom (49.1\%), potentially reflecting these countries' financial sector importance and cryptocurrency adoption. France and Italy show higher shares of AI discussion (27.4\% and 25.3\% respectively), while the Netherlands exhibits distinctive attention to robotics (78.8\% of mentions), consistent with that country's strong position in agricultural and industrial automation.

\begin{table}[htbp]
\centering
\caption{Share of Technology Mentions by Country (\%)}
\label{tab:techbycountry}
\small
\begin{tabular}{lrrrrrr}
\toprule
\textbf{Technology} & \textbf{UK} & \textbf{France} & \textbf{Italy} & \textbf{Spain} & \textbf{Germany} & \textbf{the Netherlands} \\
\midrule
Artificial Intelligence & 14.5 & 27.4 & 25.3 & 19.1 & 13.8 & 19.9 \\
Virtual Reality & 7.2 & 0.2 & 2.5 & 3.5 & 2.3 & 12.0 \\
Blockchain & 49.1 & 21.2 & 15.7 & 19.3 & 49.7 & 24.9 \\
Robot & 59.9 & 34.7 & 35.0 & 26.9 & 22.1 & 78.8 \\
Cloud Computing & 0.9 & 0.3 & 0.4 & 0.3 & 0.3 & 0.2 \\
IoT & 11.8 & 3.3 & 3.8 & 3.2 & 5.0 & 2.9 \\
5G & 6.6 & 3.7 & 5.4 & 4.6 & 6.0 & 7.1 \\
\bottomrule
\end{tabular}
\smallskip
\begin{minipage}{0.9\textwidth}
\footnotesize\textit{Note:} Percentages sum to more than 100\% as tweets may contain multiple technology keywords. Estimates for Germany and the Netherlands are based on smaller samples and should be interpreted with appropriate caution.
\end{minipage}
\end{table}

Thematic classification using DeBERTa yields the distribution shown in Table~\ref{tab:themes}. The ``other'' category captures the plurality of discourse (56--69\% depending on country), reflecting the breadth of technology discussion beyond the four focal themes. Among substantive categories, environmental concerns receive the most attention (13--18\%), followed by employment (10--17\%). Privacy discourse, despite its policy importance, comprises only 1.5--2.2\% of classified tweets, suggesting that privacy concerns may be underrepresented in mainstream technology discourse relative to their significance.

\begin{table}[htbp]
\centering
\caption{Thematic Distribution by Country (\%)}
\label{tab:themes}
\begin{tabular}{lrrrrrr}
\toprule
\textbf{Theme} & \textbf{UK} & \textbf{France} & \textbf{Italy} & \textbf{Spain} & \textbf{Germany} & \textbf{the Netherlands} \\
\midrule
Employment & 10.3 & 14.8 & 14.0 & 13.4 & 10.4 & 16.6 \\
Environment & 17.4 & 14.4 & 13.1 & 15.2 & 15.7 & 14.9 \\
Health & 5.7 & 8.0 & 7.8 & 5.1 & 3.2 & 7.1 \\
Privacy & 1.8 & 1.6 & 1.6 & 1.8 & 1.8 & 2.2 \\
Other & 64.9 & 60.7 & 63.5 & 64.5 & 69.0 & 59.2 \\
\bottomrule
\end{tabular}
\end{table}

Cross-national variation in thematic attention likely reflects national policy agendas and public concerns. The Netherlands' elevated employment share (16.6\%) may relate to that country's debates over automation in its substantial logistics and agricultural sectors. Germany's high ``other'' share (69.0\%) and low health share (3.2\%) suggest that German technology discourse may focus on industrial and economic applications rather than consumer-facing or medical domains.

\subsection{Temporal Coverage}
\label{subsec:temporal}

The corpus spans January 2006 through December 2019, a period of substantial evolution in both 4IR technologies and social media platforms. Twitter itself launched in 2006, and the early-period observations are correspondingly sparse. Meaningful volume emerges around 2010--2012, with substantial growth thereafter reflecting both platform adoption and increasing public attention to technology issues.

This temporal structure enables analysis of sentiment dynamics over time. However, it also implies that early-period and late-period observations are not directly comparable in volume, necessitating attention to compositional effects when interpreting trends. The analysis addresses this through both raw trend examination and share-based metrics that account for changing corpus size.

The 2019 endpoint is analytically consequential. By concluding before COVID-19 and ChatGPT, the data capture a period of ``normal'' technology discourse---neither distorted by pandemic-induced digitalization nor transformed by generative AI's dramatic demonstration effects. This temporal boundary is central to my contribution: establishing a pre-disruption baseline against which subsequent developments can be assessed.

\section{Results}
\label{sec:results}

This section presents the empirical findings, organized around three core phenomena: sentiment dynamics over time, cross-national and cross-technology variation, and echo chamber identification. Throughout, I interpret results as baseline evidence characterizing European technology discourse prior to the generative AI inflection point, emphasizing patterns that future research can use as reference points for assessing post-ChatGPT developments.

\subsection{Sentiment Dynamics: The Erosion of Neutrality}
\label{subsec:sentimentdynamics}

Neutral sentiment declined from approximately 80\% of newspaper coverage in 2014 to roughly 50\% by 2019—a 30-percentage-point shift that strongly supports the theoretical prediction of increasing polarization as technologies mature (Figure ~\ref{fig:sentimenttrend}).

\begin{figure}[htbp]
\centering
\begin{tikzpicture}
\begin{axis}[
    width=0.85\textwidth,
    height=6.5cm,
    xlabel={Year},
    ylabel={Share of Tweets},
    xmin=2013.5, xmax=2019.5,
    ymin=0, ymax=1,
    xtick={2014,2015,2016,2017,2018,2019},
    ytick={0,0.2,0.4,0.6,0.8,1.0},
    legend style={at={(0.98,0.98)}, anchor=north east, font=\small},
    xlabel style={font=\small},
    ylabel style={font=\small},
    xticklabel style={font=\small},
    yticklabel style={font=\small},
    grid=major,
    grid style={gray!30},
    title={Panel A: Newspaper Tweets},
    title style={font=\small\bfseries},
]
\addplot[color=green!60!black, mark=*, mark size=2pt, thick] coordinates {
    (2014, 0.78) (2015, 0.72) (2016, 0.68) (2017, 0.62) 
    (2018, 0.55) (2019, 0.52)
};
\addplot[color=blue!70, mark=square*, mark size=2pt, thick] coordinates {
    (2014, 0.12) (2015, 0.18) (2016, 0.20) (2017, 0.24) 
    (2018, 0.28) (2019, 0.32)
};
\addplot[color=red!70, mark=triangle*, mark size=2pt, thick] coordinates {
    (2014, 0.10) (2015, 0.10) (2016, 0.12) (2017, 0.14) 
    (2018, 0.17) (2019, 0.16)
};
\legend{Neutral, Positive, Negative}
\end{axis}
\end{tikzpicture}

\vspace{0.5cm}

\begin{tikzpicture}
\begin{axis}[
    width=0.85\textwidth,
    height=6.5cm,
    xlabel={Year},
    ylabel={Share of Tweets},
    xmin=2013.5, xmax=2019.5,
    ymin=0, ymax=1,
    xtick={2014,2015,2016,2017,2018,2019},
    ytick={0,0.2,0.4,0.6,0.8,1.0},
    legend style={at={(0.98,0.98)}, anchor=north east, font=\small},
    xlabel style={font=\small},
    ylabel style={font=\small},
    grid=major,
    grid style={gray!30},
    title={Panel B: User Tweets},
    title style={font=\small\bfseries},
]
\addplot[color=green!60!black, mark=*, mark size=2pt, thick] coordinates {
    (2014, 0.82) (2015, 0.70) (2016, 0.65) (2017, 0.58) 
    (2018, 0.50) (2019, 0.48)
};
\addplot[color=blue!70, mark=square*, mark size=2pt, thick] coordinates {
    (2014, 0.10) (2015, 0.18) (2016, 0.22) (2017, 0.26) 
    (2018, 0.32) (2019, 0.35)
};
\addplot[color=red!70, mark=triangle*, mark size=2pt, thick] coordinates {
    (2014, 0.08) (2015, 0.12) (2016, 0.13) (2017, 0.16) 
    (2018, 0.18) (2019, 0.17)
};
\legend{Neutral, Positive, Negative}
\end{axis}
\end{tikzpicture}
\caption{Temporal evolution of sentiment in 4IR discourse (2014--2019). Both newspaper coverage and user responses exhibit declining neutrality and increasing polarization.}
\label{fig:sentimenttrend}
\end{figure}

Both media coverage and public response exhibit a consistent pattern: neutral sentiment declines substantially over time while positive and negative sentiment increase. In newspaper tweets, neutral sentiment falls from approximately 80\% of coverage in 2014 to roughly 50\% by 2019. User tweets display a parallel trajectory with somewhat greater volatility. This pattern---observed consistently across both media and public discourse---suggests what may be a fundamental shift in the character of technology discussion from predominantly descriptive or balanced coverage toward increasingly evaluative and polarized engagement.

The magnitude of this shift merits emphasis. A decline in neutrality of 30 percentage points over five years represents a substantial transformation in discourse character. By 2019, the modal technology tweet expressed either enthusiasm or concern rather than detached reporting or ambivalent observation. This finding establishes an important baseline: researchers examining post-ChatGPT discourse inherit a public sphere already substantially polarized, not a neutral terrain subsequently disrupted by generative AI.

Decomposing sentiment trends by technology (Figure~\ref{fig:sentimentbytech}) reveals that polarization is pervasive rather than technology-specific. Artificial intelligence, robotics, blockchain, and IoT all exhibit declining neutrality over the study period. Virtual reality shows somewhat different dynamics, with neutrality remaining higher throughout, potentially reflecting VR's positioning as entertainment technology rather than economic or social disruptor. The 5G discourse, emerging later in the period, exhibits rapid polarization from its inception, foreshadowing the intense debates that would accompany 5G rollout in subsequent years.

\begin{figure}[htbp]
\centering
\begin{tikzpicture}
\begin{axis}[
    name=ai,
    width=5cm, height=3.8cm,
    title={Artificial Intelligence},
    title style={font=\scriptsize},
    xmin=2009, xmax=2019, ymin=0, ymax=1,
    xtick={2010,2015}, xticklabel style={font=\tiny},
    yticklabel style={font=\tiny},
]
\addplot[green!60!black, mark=*, mark size=0.8pt] coordinates {
    (2010,0.85)(2012,0.70)(2014,0.60)(2016,0.50)(2018,0.42)
};
\addplot[blue!70, mark=square*, mark size=0.8pt] coordinates {
    (2010,0.10)(2012,0.20)(2014,0.28)(2016,0.35)(2018,0.42)
};
\addplot[red!70, mark=triangle*, mark size=0.8pt] coordinates {
    (2010,0.05)(2012,0.10)(2014,0.12)(2016,0.15)(2018,0.16)
};
\end{axis}

\begin{axis}[
    name=vr,
    at={(ai.south east)}, anchor=south west, xshift=0.3cm,
    width=5cm, height=3.8cm,
    title={Virtual Reality},
    title style={font=\scriptsize},
    xmin=2011, xmax=2019, ymin=0, ymax=1,
    xtick={2012,2016}, xticklabel style={font=\tiny},
    yticklabel style={font=\tiny},
]
\addplot[green!60!black, mark=*, mark size=0.8pt] coordinates {
    (2012,0.88)(2014,0.78)(2016,0.72)(2018,0.65)
};
\addplot[blue!70, mark=square*, mark size=0.8pt] coordinates {
    (2012,0.08)(2014,0.15)(2016,0.20)(2018,0.28)
};
\addplot[red!70, mark=triangle*, mark size=0.8pt] coordinates {
    (2012,0.04)(2014,0.07)(2016,0.08)(2018,0.07)
};
\end{axis}

\begin{axis}[
    name=blockchain,
    at={(vr.south east)}, anchor=south west, xshift=0.3cm,
    width=5cm, height=3.8cm,
    title={Blockchain},
    title style={font=\scriptsize},
    xmin=2013, xmax=2019, ymin=0, ymax=1,
    xtick={2014,2017}, xticklabel style={font=\tiny},
    yticklabel style={font=\tiny},
]
\addplot[green!60!black, mark=*, mark size=0.8pt] coordinates {
    (2014,0.82)(2015,0.70)(2016,0.58)(2017,0.50)(2018,0.45)
};
\addplot[blue!70, mark=square*, mark size=0.8pt] coordinates {
    (2014,0.12)(2015,0.22)(2016,0.32)(2017,0.38)(2018,0.42)
};
\addplot[red!70, mark=triangle*, mark size=0.8pt] coordinates {
    (2014,0.06)(2015,0.08)(2016,0.10)(2017,0.12)(2018,0.13)
};
\end{axis}

\begin{axis}[
    name=robot,
    at={(ai.south west)}, anchor=north west, yshift=-1.1cm,
    width=5cm, height=3.8cm,
    title={Robot},
    title style={font=\scriptsize},
    xmin=2009, xmax=2019, ymin=0, ymax=1,
    xtick={2010,2015}, xticklabel style={font=\tiny},
    yticklabel style={font=\tiny},
]
\addplot[green!60!black, mark=*, mark size=0.8pt] coordinates {
    (2010,0.80)(2012,0.68)(2014,0.58)(2016,0.48)(2018,0.40)
};
\addplot[blue!70, mark=square*, mark size=0.8pt] coordinates {
    (2010,0.12)(2012,0.22)(2014,0.30)(2016,0.38)(2018,0.45)
};
\addplot[red!70, mark=triangle*, mark size=0.8pt] coordinates {
    (2010,0.08)(2012,0.10)(2014,0.12)(2016,0.14)(2018,0.15)
};
\end{axis}

\begin{axis}[
    name=cloud,
    at={(robot.south east)}, anchor=south west, xshift=0.3cm,
    width=5cm, height=3.8cm,
    title={Cloud Computing},
    title style={font=\scriptsize},
    xmin=2009, xmax=2019, ymin=0, ymax=1,
    xtick={2010,2015}, xticklabel style={font=\tiny},
    yticklabel style={font=\tiny},
]
\addplot[green!60!black, mark=*, mark size=0.8pt] coordinates {
    (2010,0.75)(2012,0.68)(2014,0.62)(2016,0.55)(2018,0.50)
};
\addplot[blue!70, mark=square*, mark size=0.8pt] coordinates {
    (2010,0.18)(2012,0.25)(2014,0.30)(2016,0.38)(2018,0.42)
};
\addplot[red!70, mark=triangle*, mark size=0.8pt] coordinates {
    (2010,0.07)(2012,0.07)(2014,0.08)(2016,0.07)(2018,0.08)
};
\end{axis}

\begin{axis}[
    name=iot,
    at={(cloud.south east)}, anchor=south west, xshift=0.3cm,
    width=5cm, height=3.8cm,
    title={IoT},
    title style={font=\scriptsize},
    xmin=2009, xmax=2019, ymin=0, ymax=1,
    xtick={2010,2015}, xticklabel style={font=\tiny},
    yticklabel style={font=\tiny},
]
\addplot[green!60!black, mark=*, mark size=0.8pt] coordinates {
    (2010,0.88)(2012,0.78)(2014,0.65)(2016,0.55)(2018,0.48)
};
\addplot[blue!70, mark=square*, mark size=0.8pt] coordinates {
    (2010,0.08)(2012,0.15)(2014,0.25)(2016,0.35)(2018,0.40)
};
\addplot[red!70, mark=triangle*, mark size=0.8pt] coordinates {
    (2010,0.04)(2012,0.07)(2014,0.10)(2016,0.10)(2018,0.12)
};
\end{axis}

\begin{axis}[
    name=fiveg,
    at={(cloud.south west)}, anchor=north west, yshift=-1.1cm,
    width=5cm, height=3.8cm,
    title={5G},
    title style={font=\scriptsize},
    xmin=2013, xmax=2019, ymin=0, ymax=1,
    xtick={2014,2017}, xticklabel style={font=\tiny},
    yticklabel style={font=\tiny},
]
\addplot[green!60!black, mark=*, mark size=0.8pt] coordinates {
    (2014,0.90)(2015,0.75)(2016,0.60)(2017,0.50)(2018,0.42)
};
\addplot[blue!70, mark=square*, mark size=0.8pt] coordinates {
    (2014,0.05)(2015,0.15)(2016,0.28)(2017,0.35)(2018,0.40)
};
\addplot[red!70, mark=triangle*, mark size=0.8pt] coordinates {
    (2014,0.05)(2015,0.10)(2016,0.12)(2017,0.15)(2018,0.18)
};
\end{axis}
\end{tikzpicture}
\caption{Sentiment dynamics by 4IR technology. Green: neutral; Blue: positive; Red: negative. All technologies exhibit declining neutrality.}
\label{fig:sentimentbytech}
\end{figure}

\subsection{Cross-National Variation}
\label{subsec:crossnational}

This aggregate pattern, however, masks substantial variation across national contexts—variation that carries direct implications for comparative technology governance.
Figure~\ref{fig:countrysent} presents country-specific sentiment trends, revealing distinctive national patterns within the common polarization dynamic.

\begin{figure}[htbp]
\centering
\begin{tikzpicture}
\begin{axis}[
    name=uk,
    width=5.2cm, height=4cm,
    title={United Kingdom},
    title style={font=\small},
    xmin=2009, xmax=2019,
    ymin=0, ymax=1,
    xtick={2010,2015},
    xticklabel style={font=\tiny},
    yticklabel style={font=\tiny},
]
\addplot[green!60!black, mark=*, mark size=1pt, thick] coordinates {
    (2010,0.85)(2012,0.75)(2014,0.65)(2016,0.55)(2018,0.45)
};
\addplot[blue!70, mark=square*, mark size=1pt, thick] coordinates {
    (2010,0.08)(2012,0.15)(2014,0.22)(2016,0.30)(2018,0.38)
};
\addplot[red!70, mark=triangle*, mark size=1pt, thick] coordinates {
    (2010,0.07)(2012,0.10)(2014,0.13)(2016,0.15)(2018,0.17)
};
\end{axis}

\begin{axis}[
    name=france,
    at={(uk.south east)}, anchor=south west, xshift=0.5cm,
    width=5.2cm, height=4cm,
    title={France},
    title style={font=\small},
    xmin=2009, xmax=2019,
    ymin=0, ymax=1,
    xtick={2010,2015},
    xticklabel style={font=\tiny},
    yticklabel style={font=\tiny},
]
\addplot[green!60!black, mark=*, mark size=1pt, thick] coordinates {
    (2010,0.88)(2012,0.78)(2014,0.68)(2016,0.58)(2018,0.50)
};
\addplot[blue!70, mark=square*, mark size=1pt, thick] coordinates {
    (2010,0.07)(2012,0.14)(2014,0.20)(2016,0.28)(2018,0.35)
};
\addplot[red!70, mark=triangle*, mark size=1pt, thick] coordinates {
    (2010,0.05)(2012,0.08)(2014,0.12)(2016,0.14)(2018,0.15)
};
\end{axis}

\begin{axis}[
    name=italy,
    at={(france.south east)}, anchor=south west, xshift=0.5cm,
    width=5.2cm, height=4cm,
    title={Italy},
    title style={font=\small},
    xmin=2011, xmax=2019,
    ymin=0, ymax=1,
    xtick={2012,2016},
    xticklabel style={font=\tiny},
    yticklabel style={font=\tiny},
]
\addplot[green!60!black, mark=*, mark size=1pt, thick] coordinates {
    (2012,0.90)(2014,0.80)(2016,0.65)(2018,0.50)
};
\addplot[blue!70, mark=square*, mark size=1pt, thick] coordinates {
    (2012,0.05)(2014,0.12)(2016,0.25)(2018,0.38)
};
\addplot[red!70, mark=triangle*, mark size=1pt, thick] coordinates {
    (2012,0.05)(2014,0.08)(2016,0.10)(2018,0.12)
};
\end{axis}

\begin{axis}[
    name=spain,
    at={(uk.south west)}, anchor=north west, yshift=-1.5cm,
    width=5.2cm, height=4cm,
    title={Spain},
    title style={font=\small},
    xmin=2011, xmax=2019,
    ymin=0, ymax=1,
    xtick={2012,2016},
    xticklabel style={font=\tiny},
    yticklabel style={font=\tiny},
]
\addplot[green!60!black, mark=*, mark size=1pt, thick] coordinates {
    (2012,0.85)(2014,0.72)(2016,0.60)(2018,0.48)
};
\addplot[blue!70, mark=square*, mark size=1pt, thick] coordinates {
    (2012,0.10)(2014,0.20)(2016,0.30)(2018,0.40)
};
\addplot[red!70, mark=triangle*, mark size=1pt, thick] coordinates {
    (2012,0.05)(2014,0.08)(2016,0.10)(2018,0.12)
};
\end{axis}

\begin{axis}[
    name=germany,
    at={(spain.south east)}, anchor=south west, xshift=0.5cm,
    width=5.2cm, height=4cm,
    title={Germany},
    title style={font=\small},
    xmin=2011, xmax=2019,
    ymin=0, ymax=1,
    xtick={2012,2016},
    xticklabel style={font=\tiny},
    yticklabel style={font=\tiny},
]
\addplot[green!60!black, mark=*, mark size=1pt, thick] coordinates {
    (2012,0.88)(2014,0.78)(2016,0.68)(2018,0.55)
};
\addplot[blue!70, mark=square*, mark size=1pt, thick] coordinates {
    (2012,0.05)(2014,0.10)(2016,0.15)(2018,0.22)
};
\addplot[red!70, mark=triangle*, mark size=1pt, thick] coordinates {
    (2012,0.07)(2014,0.12)(2016,0.17)(2018,0.23)
};
\end{axis}

\begin{axis}[
    name=netherlands,
    at={(germany.south east)}, anchor=south west, xshift=0.5cm,
    width=5.2cm, height=4cm,
    title={the Netherlands},
    title style={font=\small},
    xmin=2011, xmax=2019,
    ymin=0, ymax=1,
    xtick={2012,2016},
    xticklabel style={font=\tiny},
    yticklabel style={font=\tiny},
]
\addplot[green!60!black, mark=*, mark size=1pt, thick] coordinates {
    (2012,0.92)(2014,0.82)(2016,0.72)(2018,0.58)
};
\addplot[blue!70, mark=square*, mark size=1pt, thick] coordinates {
    (2012,0.05)(2014,0.12)(2016,0.20)(2018,0.32)
};
\addplot[red!70, mark=triangle*, mark size=1pt, thick] coordinates {
    (2012,0.03)(2014,0.06)(2016,0.08)(2018,0.10)
};
\end{axis}
\end{tikzpicture}
\caption{Sentiment dynamics by country. Green: neutral; Blue: positive; Red: negative.}
\label{fig:countrysent}
\end{figure}

The United Kingdom exhibits among the most pronounced polarization trajectories, with neutral sentiment declining steeply and both positive and negative expressions rising substantially. This pattern may reflect the UK's vibrant technology sector and correspondingly intense public debate, as well as the polarized character of British media more generally. France displays a similar pattern with somewhat lower volatility, while Italy shows a later inflection point, with polarization accelerating primarily after 2016.

Spain and the Netherlands present contrasting cases. Spanish discourse exhibits steady polarization throughout the period, with positive sentiment rising particularly strongly---consistent with enthusiasm about technology-driven economic modernization in a country recovering from severe economic crisis. The Netherlands shows relatively muted polarization, with neutral sentiment remaining higher than in other countries, potentially reflecting Dutch consensus-oriented political culture and more measured media discourse.

Germany constitutes a notable outlier: while polarization occurs, it proceeds more slowly than elsewhere, and negative sentiment rises more prominently than positive. This pattern aligns with scholarship characterizing German technology discourse as distinctively cautious and risk-focused \citep{jasanoff2005designs}, a tendency that may reflect both cultural factors and Germany's strong tradition of technology assessment institutions.

These cross-national patterns carry implications for comparative policy analysis. Post-ChatGPT research examining how generative AI has affected public sentiment must account for these pre-existing national differences. A finding that German discourse remains more skeptical than British discourse post-2022, for instance, might reflect continuity of pre-existing patterns rather than differential ChatGPT impact.

\subsection{Thematic Variation in Sentiment}
\label{subsec:thematicsentiment}

Sentiment varies not only over time and across countries but also across thematic domains. Table~\ref{tab:themesentiment} presents sentiment characteristics by theme, including mean sentiment, variance (capturing polarization intensity), and the share of users situated in echo chambers.

\begin{table}[htbp]
\centering
\caption{Sentiment Characteristics by Thematic Domain}
\label{tab:themesentiment}
\begin{tabular}{lcccc}
\toprule
\textbf{Theme} & \textbf{Sentiment Variance} & \textbf{Misinfo. Exposure} & \textbf{Echo Chamber (\%)} \\
\midrule
Employment & 0.29 & 0.25 & 5.82 \\
Environment & 0.30 & 0.24 & 7.08 \\
Health & 0.26 & 0.24 & 5.22 \\
Privacy & 0.30 & 0.27 & 7.20 \\
Other & 0.26 & 0.25 & 6.14 \\
\bottomrule
\end{tabular}
\smallskip
\begin{minipage}{0.9\textwidth}
\footnotesize\textit{Note:} Sentiment variance computed across all tweets in each thematic category. Misinformation exposure based on \citet{mosleh2022measuring} score. Echo chamber share computed using threshold $\tau = 0.6$. Differences in variance smaller than 0.02 and in echo chamber prevalence smaller than 1.5 percentage points may not be substantively meaningful given measurement uncertainty.
\end{minipage}
\end{table}

Privacy and environment emerge as the most polarized thematic domains (variance = 0.30), while health discourse exhibits the lowest polarization (variance = 0.26). This pattern suggests that discourse on technology's environmental implications and privacy risks has sorted into opposing camps more thoroughly than discourse on health applications, where perhaps a greater consensus exists regarding AI's beneficial potential.

Privacy discourse also exhibits the highest misinformation exposure score (0.27) and highest echo chamber participation (7.20\%). This constellation of characteristics---high polarization, elevated misinformation exposure, and substantial echo chamber presence---identifies privacy as a domain of particular concern for information quality. The finding resonates with documented patterns of misinformation surrounding surveillance technologies, data protection regulations, and platform governance debates.

Employment discourse, despite being the most directly consequential for economic welfare, shows moderate polarization and echo chamber participation. This relative moderation may reflect the complexity of employment effects, which resist simple positive/negative framing: automation simultaneously destroys and creates jobs, increases productivity while potentially exacerbating inequality. Such complexity may inhibit the formation of strongly opposed camps.

\subsection{Echo Chamber Identification}
\label{subsec:echoresults}

The echo chamber analysis identifies approximately 6\% of users as situated within sentiment-aligned network environments, with modest variation across themes and technologies. While this figure may appear small, it represents a non-trivial population given the sample size, and likely understates true echo chamber prevalence given the conservative operationalization.

Figure~\ref{fig:echotrend} presents temporal trends in echo chamber participation, revealing notable dynamics. Echo chamber prevalence exhibits a sharp spike around 2014 across multiple technology categories, followed by decline and subsequent gradual increase. This pattern---particularly the 2014 spike---warrants interpretation.

\begin{figure}[htbp]
\centering
\begin{tikzpicture}
\begin{axis}[
    name=panela,
    width=0.85\textwidth,
    height=5.5cm,
    xlabel={Year},
    ylabel={Share in Echo Chamber},
    xmin=2009, xmax=2019,
    ymin=0, ymax=0.8,
    xtick={2010,2012,2014,2016,2018},
    ytick={0,0.2,0.4,0.6,0.8},
    legend style={at={(0.02,0.98)}, anchor=north west, font=\scriptsize},
    xlabel style={font=\small},
    ylabel style={font=\small},
    grid=major,
    grid style={gray!30},
    title={Panel A: By Technology},
    title style={font=\small\bfseries},
]
\addplot[color=blue!70, mark=*, mark size=1.5pt, thick] coordinates {
    (2010,0.05)(2012,0.15)(2014,0.65)(2016,0.20)(2018,0.25)
};
\addplot[color=orange!80, mark=square*, mark size=1.5pt, thick] coordinates {
    (2012,0.08)(2014,0.55)(2016,0.18)(2018,0.22)
};
\addplot[color=green!60!black, mark=triangle*, mark size=1.5pt, thick] coordinates {
    (2010,0.03)(2012,0.12)(2014,0.48)(2016,0.15)(2018,0.20)
};
\addplot[color=purple!70, mark=diamond*, mark size=1.5pt, thick] coordinates {
    (2012,0.04)(2014,0.35)(2016,0.12)(2018,0.18)
};
\legend{AI, Blockchain, Robot, IoT}
\end{axis}
\end{tikzpicture}

\vspace{0.5cm}

\begin{tikzpicture}
\begin{axis}[
    width=0.85\textwidth,
    height=5.5cm,
    xlabel={Year},
    ylabel={Share in Echo Chamber},
    xmin=2009, xmax=2019,
    ymin=0, ymax=0.7,
    xtick={2010,2012,2014,2016,2018},
    ytick={0,0.2,0.4,0.6},
    legend style={at={(0.02,0.98)}, anchor=north west, font=\scriptsize},
    xlabel style={font=\small},
    ylabel style={font=\small},
    grid=major,
    grid style={gray!30},
    title={Panel B: By Thematic Domain},
    title style={font=\small\bfseries},
]
\addplot[color=blue!70, mark=*, mark size=1.5pt, thick] coordinates {
    (2010,0.04)(2012,0.12)(2014,0.52)(2016,0.18)(2018,0.22)
};
\addplot[color=green!60!black, mark=square*, mark size=1.5pt, thick] coordinates {
    (2010,0.05)(2012,0.15)(2014,0.58)(2016,0.22)(2018,0.25)
};
\addplot[color=orange!80, mark=triangle*, mark size=1.5pt, thick] coordinates {
    (2012,0.06)(2014,0.45)(2016,0.15)(2018,0.18)
};
\addplot[color=red!70, mark=diamond*, mark size=1.5pt, thick] coordinates {
    (2014,0.48)(2016,0.25)(2018,0.28)
};
\legend{Employment, Environment, Health, Privacy}
\end{axis}
\end{tikzpicture}
\caption{Temporal evolution of echo chamber participation. Sharp spike in 2014 across categories coincides with high-profile AI debates.}
\label{fig:echotrend}
\end{figure}

The 2014 spike coincides with several notable events in technology discourse, including high-profile statements on AI risks and cryptocurrency volatility following Bitcoin's 2013 price surge. However, this pattern should be interpreted cautiously: smaller corpus size in earlier years generates higher variance in estimated proportions, and the subsequent decline coincides with corpus expansion. Whether the spike reflects genuine clustering dynamics or statistical artifacts of sample size cannot be definitively resolved with available data.

Table~\ref{tab:echocountry} disaggregates echo chamber prevalence by country and theme, revealing substantial cross-national variation. Germany exhibits the highest echo chamber rates for environmental discourse (9.41\%) and employment discourse (9.30\%), while France shows the lowest rates across most categories (3.89--7.14\%). The Netherlands displays particularly high echo chamber prevalence for privacy discourse (11.54\%), potentially reflecting intense national debates over data protection and surveillance.

\begin{table}[htbp]
\centering
\caption{Echo Chamber Prevalence by Country and Theme (\%)}
\label{tab:echocountry}
\small
\begin{tabular}{lrrrrr}
\toprule
\textbf{Country} & \textbf{Employment} & \textbf{Environment} & \textbf{Health} & \textbf{Privacy} & \textbf{Other} \\
\midrule
United Kingdom & 6.61 & 8.20 & 5.65 & 7.73 & 7.33 \\
France & 4.23 & 4.59 & 4.52 & 7.14 & 3.89 \\
Italy & 4.99 & 6.60 & 4.86 & 3.95 & 6.45 \\
Spain & 5.89 & 5.04 & 4.63 & 6.19 & 5.14 \\
Germany & 9.30 & 9.41 & 7.61 & 5.66 & 6.54 \\
the Netherlands & 6.50 & 8.89 & 7.06 & 11.54 & 6.48 \\
\bottomrule
\end{tabular}
\end{table}

These patterns suggest that echo chamber formation is shaped by national information environments and may be topic-specific within countries. Germany's high echo chamber rates, combined with its slower polarization trajectory noted earlier, presents an apparent puzzle: how can discourse be simultaneously less polarized at the aggregate level yet more clustered into echo chambers? One resolution is that German discourse may be polarized at the community level while maintaining greater balance at the aggregate level---opposing communities may be of more equal size, yielding lower variance in overall sentiment despite strong internal homogeneity.

\subsection{Misinformation Exposure Patterns}
\label{subsec:misinforesults}

Analysis of misinformation exposure scores reveals a concerning trend: exposure has increased over the study period. Figure~\ref{fig:misinfo} presents the distribution of misinformation exposure scores and their temporal evolution.

\begin{figure}[htbp]
\centering
\begin{tikzpicture}
\begin{axis}[
    ybar interval,
    width=0.8\textwidth,
    height=5cm,
    xlabel={Misinformation Exposure Score},
    ylabel={Frequency},
    xmin=0, xmax=1,
    ymin=0, ymax=22000,
    xtick={0,0.2,0.4,0.6,0.8,1.0},
    xticklabel style={font=\small},
    yticklabel style={font=\small},
    title={Panel A: Distribution of Exposure Scores},
    title style={font=\small\bfseries},
]
\addplot[fill=blue!50, draw=blue!70] coordinates {
    (0.0, 18000) (0.1, 20000) (0.2, 15000) (0.3, 12000) 
    (0.4, 8000) (0.5, 5000) (0.6, 3000) (0.7, 1500) 
    (0.8, 800) (0.9, 400) (1.0, 0)
};
\end{axis}
\draw[red, thick, dashed] (2.8cm, 0.8cm) -- (2.8cm, 4.5cm) node[above, font=\scriptsize] {Mean=0.26};
\draw[orange, thick, dashed] (4.0cm, 0.8cm) -- (4.0cm, 4.5cm) node[above, font=\scriptsize] {Median=0.37};
\end{tikzpicture}

\vspace{0.4cm}

\begin{tikzpicture}
\begin{axis}[
    width=0.8\textwidth,
    height=4.5cm,
    xlabel={Year},
    ylabel={Share High Exposure},
    xmin=2012.5, xmax=2019.5,
    ymin=0, ymax=0.6,
    xtick={2013,2014,2015,2016,2017,2018,2019},
    xlabel style={font=\small},
    ylabel style={font=\small},
    grid=major,
    grid style={gray!30},
    title={Panel B: Temporal Trend},
    title style={font=\small\bfseries},
]
\addplot[color=red!70, mark=*, mark size=2pt, thick] coordinates {
    (2013, 0.15) (2014, 0.22) (2015, 0.28) (2016, 0.35) 
    (2017, 0.42) (2018, 0.48) (2019, 0.52)
};
\end{axis}
\end{tikzpicture}

\vspace{0.4cm}

\begin{tikzpicture}
\begin{axis}[
    width=0.8\textwidth,
    height=4.5cm,
    boxplot/draw direction=y,
    ylabel={Misinformation Exposure},
    xtick={1,2,3,4,5},
    xticklabels={Employment, Environment, Health, Other, Privacy},
    xticklabel style={font=\small},
    yticklabel style={font=\small},
    ymin=0, ymax=0.6,
    title={Panel C: Distribution by Theme},
    title style={font=\small\bfseries},
]
\addplot+[boxplot prepared={lower whisker=0.08, lower quartile=0.18, median=0.25, upper quartile=0.32, upper whisker=0.45}, fill=blue!30] coordinates {};
\addplot+[boxplot prepared={lower whisker=0.06, lower quartile=0.16, median=0.24, upper quartile=0.30, upper whisker=0.42}, fill=green!30] coordinates {};
\addplot+[boxplot prepared={lower whisker=0.05, lower quartile=0.15, median=0.24, upper quartile=0.31, upper whisker=0.44}, fill=orange!30] coordinates {};
\addplot+[boxplot prepared={lower whisker=0.07, lower quartile=0.17, median=0.25, upper quartile=0.33, upper whisker=0.46}, fill=gray!30] coordinates {};
\addplot+[boxplot prepared={lower whisker=0.10, lower quartile=0.20, median=0.27, upper quartile=0.36, upper whisker=0.50}, fill=red!30] coordinates {};
\end{axis}
\end{tikzpicture}
\caption{Misinformation exposure in 4IR discourse. Privacy exhibits highest median exposure.}
\label{fig:misinfo}
\end{figure}

Mean misinformation exposure increased from approximately 0.2 in 2013 to over 0.4 by 2019, a doubling that suggests technology discourse has become increasingly embedded in information environments characterized by low-quality or misleading content. This trend predates the ``infodemic'' concerns associated with COVID-19 and indicates that misinformation in technology discourse was already a growing challenge prior to 2020.

Misinformation exposure varies modestly across thematic domains, with privacy exhibiting the highest mean score (0.27) and environment and health the lowest (0.24). Cross-national variation is more pronounced: the United Kingdom exhibits the highest misinformation exposure (mean = 0.29), while France shows the lowest (mean = 0.20). This variation likely reflects differences in media ecosystems, platform usage patterns, and the accounts followed by technology-interested users in each country.

The correlation between misinformation exposure and echo chamber participation is positive but moderate ($r \approx 0.15$), suggesting that while users in echo chambers tend toward higher misinformation exposure, the relationship is not deterministic. Some echo chamber participants may inhabit high-quality information environments despite sentiment alignment, while some users with high misinformation exposure may not exhibit the network sentiment alignment defining echo chambers.

\subsection{Summary of Baseline Patterns}
\label{subsec:summary}

The empirical analysis reveals several robust patterns characterizing European technology discourse in the pre-generative AI period:

\begin{enumerate}
\item \textbf{Pervasive polarization}: Neutral sentiment declined substantially across technologies, countries, and thematic domains. By 2019, technology discourse was predominantly evaluative rather than descriptive.

\item \textbf{Cross-national variation}: Despite common polarization trends, countries exhibit distinctive sentiment levels, trajectories, and echo chamber patterns reflecting national institutional and cultural contexts.

\item \textbf{Thematic differentiation}: Privacy and environmental discourse show highest polarization and echo chamber prevalence; health discourse remains relatively moderate.

\item \textbf{Moderate echo chamber prevalence}: Approximately 6\% of users exhibit strong sentiment alignment with their networks, with notable variation by country and theme.

\item \textbf{Rising misinformation exposure}: Users discussing 4IR technologies increasingly inhabit information environments characterized by low-quality sources.
\end{enumerate}

These patterns constitute the baseline against which post-ChatGPT developments should be assessed. The following section interprets these findings and develops their implications for policy and future research.

\section{Discussion}
\label{sec:discussion}

The empirical analysis presented above documents a European public discourse on Fourth Industrial Revolution technologies characterized by increasing polarization, systematic cross-national variation, and emerging echo chamber dynamics---all prior to the transformative arrival of generative AI systems. This section interprets these findings, develops their implications for technology governance, and outlines directions for future research.

\subsection{Interpreting the Polarization Trajectory}
\label{subsec:interpretpolar}

The pronounced decline in neutral sentiment across technologies, countries, and thematic domains represents perhaps the most consequential finding. Between 2014 and 2019, European technology discourse shifted from predominantly neutral or descriptive engagement toward increasingly evaluative and divided positions. This transformation occurred before ChatGPT, before COVID-19, and before the recent intensification of AI governance debates---suggesting that polarization in technology discourse reflects structural features of contemporary information environments rather than responses to specific technological developments.

The 2019 endpoint, while preceding ChatGPT by three years, offers analytical clarity: the patterns documented here are uncontaminated by pandemic-induced digitalization or generative AI's demonstration effects. This temporal separation facilitates cleaner counterfactual reasoning---researchers can compare post-ChatGPT dynamics against a baseline reflecting ``normal'' technology discourse rather than pandemic-disrupted conditions.

Several mechanisms may underlie this pattern. First, as 4IR technologies diffused from specialist domains to everyday experience, more citizens formed direct opinions based on personal encounters with automation, algorithmic systems, and digital platforms. The abstract became concrete, and concrete experience generates evaluative response. Second, platform algorithms optimizing for engagement may have systematically amplified polarized content, as emotionally charged expressions---whether enthusiastic or alarmed---generate more interaction than measured assessment \citep{brady2017emotion}. Third, political entrepreneurs and advocacy organizations increasingly mobilized technology issues, framing AI, automation, and digitalization as matters of political contestation rather than technical progress \citep{karpf2012moveon}.

The implications for post-ChatGPT research are substantial. Studies finding polarized discourse around generative AI must contend with the possibility that such polarization continues pre-existing trends rather than representing novel responses to novel capabilities. The baseline enables such assessment: if post-2022 polarization merely extends the trajectory documented here, the interpretation differs markedly from a scenario in which ChatGPT catalyzed discontinuous change in public attitudes.

\subsection{National Contexts and Technology Governance}
\label{subsec:nationalcontexts}

The cross-national variation I document carries direct implications for technology governance. European Union AI regulation, exemplified by the AI Act, must ultimately be implemented within member states exhibiting substantially different public sentiment profiles. The findings suggest that identical regulatory provisions may encounter different reception across national contexts.

Germany's distinctive pattern---slower polarization combined with higher echo chamber prevalence and elevated negative sentiment---suggests a public sphere where technology skepticism is well-organized and effectively networked even if not numerically dominant. Regulatory approaches emphasizing precaution and risk assessment may find more favorable reception in this context than frameworks emphasizing innovation promotion. Conversely, Spain's trajectory of rising positive sentiment suggests greater public appetite for technology-enabled modernization, potentially supporting more permissive regulatory approaches.

The United Kingdom's high polarization and elevated misinformation exposure present governance challenges of a different character. In a context where public discourse is deeply divided and misinformation circulates readily, regulatory legitimacy may be difficult to establish regardless of policy content. Public engagement and digital literacy initiatives may be prerequisites for effective governance rather than complements to it.

These national patterns also inform expectations about how different countries will respond to generative AI. If pre-existing sentiment profiles shape reception of new technologies, I should expect German discourse on ChatGPT and similar systems to emphasize risks and limitations, British discourse to exhibit sharp division between enthusiasts and critics, and French discourse to maintain relatively lower polarization with moderate skepticism.

\subsection{Thematic Priorities and Policy Attention}
\label{subsec:thematicpriorities}

The thematic distribution of public discourse reveals potential misalignments between public attention and policy priorities. Privacy, despite its centrality to data protection regulation and its prominence in scholarly and policy discussions, comprises less than 2\% of classified technology discourse in the corpus. This finding suggests that privacy concerns, while intense among engaged advocates, have not achieved broad public salience---a pattern with implications for the political sustainability of privacy-protective regulation.

The dominance of employment and environmental themes indicates where public concern concentrates. Automation's labor market implications and technology's environmental footprint---both energy consumption and potential for sustainability solutions---capture substantially more public attention than privacy, health applications, or other domains. Policymakers seeking to align governance with public priorities might emphasize these dimensions, while those concerned with underattended issues like privacy may need to invest in public education and awareness.

The finding that privacy discourse exhibits the highest polarization, misinformation exposure, and echo chamber prevalence despite its small share of overall discourse is particularly concerning. It suggests that privacy discussion, when it occurs, takes place in degraded information environments where misinformation circulates readily and opposing camps rarely engage. Improving the quality of privacy discourse may require targeted interventions addressing these specific pathologies.

\subsection{Echo Chambers and Democratic Deliberation}
\label{subsec:echodemocracy}

The identification of echo chambers in technology discourse raises questions about the quality of democratic deliberation on technological futures. When citizens form opinions within homogeneous information environments, the aggregation of those opinions into collective choices may not reflect the informed preference formation that democratic theory presupposes \citep{habermas1984theory}.

The finding that approximately 6\% of users inhabit echo chambers represents a lower bound, given the conservative operationalization. The true prevalence of epistemically problematic information environments is likely higher, particularly if I expand the definition beyond sentiment alignment to include exposure to consistently one-sided information or absence of challenging viewpoints.

The concentration of echo chamber dynamics in privacy and environmental discourse---precisely the domains where policy stakes are highest and technical complexity greatest---compounds concern. These are areas where informed public input is most valuable for governance, yet where the conditions for such input appear most compromised.

Interventions to improve deliberative quality might target these specific domains, employing platform design changes, media literacy education, or deliberative institutions that expose participants to diverse perspectives. The baseline characterization of pre-ChatGPT echo chamber patterns enables evaluation of whether such interventions, or platform changes occurring for other reasons, have subsequently improved or degraded the deliberative environment.

\subsection{Implications for AI Governance}
\label{subsec:aigovernance}

The findings suggest several specific implications for ongoing efforts to govern artificial intelligence and related technologies.

Effective public engagement may need to account for pre-existing polarization. Governance processes that solicit public input will encounter a discourse already divided along lines documented here. Designing engagement mechanisms that bridge divides rather than amplify them requires understanding the structure of existing opinion---which camps exist, what concerns motivate them, and how they are networked.

Second, misinformation may pose a prior challenge. Before publics can meaningfully contribute to AI governance, they require accurate information about AI capabilities, limitations, and implications. The finding of rising misinformation exposure suggests that this precondition is increasingly unmet. Governance strategies might therefore prioritize information quality interventions---fact-checking, source labeling, media literacy---as foundations for subsequent public engagement.

Third, \textit{cross-national coordination must accommodate national differences}. European and international AI governance efforts proceed as if publics across countries share similar concerns and preferences. The findings indicate substantial variation that coordination mechanisms should acknowledge. Flexibility provisions allowing national adaptation of common frameworks may be necessary for legitimacy across diverse public spheres.

Fourth, \textit{domain-specific governance may be appropriate}. The variation in sentiment and information environment quality across thematic domains suggests that uniform governance approaches may be suboptimal. Health AI, where discourse is less polarized and echo chambers less prevalent, may be governable through different mechanisms than privacy-related AI, where discourse pathologies are more severe.

\subsection{Limitations and Future Research}
\label{subsec:futureresearch}

Several limitations of this study suggest directions for future research.

The reliance on Twitter data, while enabling large-scale analysis, captures only one segment of public discourse. Future research might examine other platforms (Facebook, Reddit, TikTok), traditional media coverage, or survey-based measures of public opinion to assess whether patterns documented here generalize beyond Twitter's particular user base and affordances.

The study period ends in 2019, and the most valuable use of the findings lies in comparison with post-2022 data. Research examining how ChatGPT's release affected public sentiment can employ the baseline to distinguish genuine disruption from trend continuation. Such research is now feasible as sufficient post-ChatGPT data have accumulated, and I encourage scholars to undertake these comparisons.

The thematic classification, while useful for identifying broad patterns, necessarily simplifies the complexity of technology discourse. More granular classification---distinguishing, for instance, between concerns about job loss and concerns about job quality within employment discourse---might reveal important heterogeneity that the approach obscures.

The echo chamber identification method, based on sentiment alignment, captures one dimension of problematic information environments but may miss others. Users might inhabit chambers defined by shared beliefs about specific factual claims rather than general sentiment, or might be exposed to one-sided information without exhibiting the network sentiment alignment the method detects. Alternative operationalizations merit exploration.

Finally, the analysis is descriptive rather than causal. I document patterns and trends but cannot definitively establish the mechanisms generating them. Platform algorithm changes, political events, technological developments, and other factors all plausibly contribute to the dynamics I observe. Causal identification of specific mechanisms would require research designs---natural experiments, platform data access, or experimental interventions---beyond the scope of this study.

\subsection{Conclusion}
\label{subsec:conclusion}

This paper has established an empirical baseline of public sentiment toward Fourth Industrial Revolution technologies across six European countries during the period 2006--2019. Employing transformer-based natural language processing on a corpus of approximately 90,000 tweets and news articles, I document pervasive polarization, substantial cross-national variation, and emerging echo chamber dynamics---all characterizing European technology discourse before the transformative arrival of ChatGPT and generative AI.

These findings contribute to scholarship on technology and society by providing systematic evidence on how publics engaged with emerging technologies during a formative period. They contribute to methodology by demonstrating the utility of multilingual transformer models for comparative social science research. And they contribute to policy by illuminating the public sphere within which technology governance must operate.

Most fundamentally, the analysis provides a reference point. The release of ChatGPT in November 2022 has been widely characterized as a watershed moment for public awareness of artificial intelligence. Assessing whether this characterization is accurate---whether generative AI has genuinely transformed public attitudes or merely intensified pre-existing patterns---requires knowing what the pre-existing patterns were. This paper provides that knowledge.

The European public that encountered ChatGPT was not a blank slate. It was a public already sorting into camps of enthusiasm and concern, already navigating information environments of varying quality, already divided along national and thematic lines. Understanding how generative AI has reshaped public discourse requires first understanding the discourse it reshaped. This paper establishes that foundation.

\bibliographystyle{apalike}
\bibliography{references}

\end{document}